\documentclass[11pt]{article}
\pdfoutput=1
\usepackage[latin1]{inputenc}
\usepackage{amsmath}
\usepackage{amsfonts}
\usepackage{amssymb}
\usepackage{pstricks}
\usepackage{amsthm}
\usepackage{mathrsfs}
\usepackage{amssymb}
\usepackage{cancel}
\usepackage{slashed}
\usepackage{graphicx}
\usepackage[font=small,labelfont=bf]{caption}
\usepackage{float}
\usepackage{csquotes}
\usepackage[bottom]{footmisc} 
\usepackage{tikz}
\usepackage{setspace}
\usepackage{bbding}
\usepackage{cite}
\usetikzlibrary{matrix,arrows,decorations.pathmorphing}
\usepackage{framed}
\usepackage{color}
\usepackage{wrapfig}
\definecolor{shadecolor}{RGB}{224,238,238}
\makeatletter

\@addtoreset{equation}{section}
\makeatother

\def\lsim{\;\raise0.3ex\hbox{$<$\kern-0.75em\raise-1.1ex\hbox{$\sim$}}\;}
\def\gsim{\;\raise0.3ex\hbox{$>$\kern-0.75em\raise-1.1ex\hbox{$\sim$}}\;}
\def\beq{\begin{equation}}   \def\eeq{\end{equation}}
\def\ba{\begin{array}}       \def\ea{\end{array}}
\def\bea{\begin{eqnarray}}   \def\eea{\end{eqnarray}}

\def\nl{\newline}

\theoremstyle{definition} 
\date{\today}

\begin{document}

\begin{titlepage}
\begin{flushright}
LPT Orsay 16-91
\end{flushright}


\begin{center}

\begin{doublespace}

\vspace{1cm}
{\Large\bf Present Status and Future Tests of the Higgsino-Singlino Sector in the NMSSM} \\
\vspace{2cm}

{\bf{Ulrich Ellwanger$^{a,b}$}}\\
\vspace{1cm}
{\it  $^a$ Laboratoire de Physique Th\'eorique, UMR 8627, CNRS, Universit\'e de Paris-Sud, Universit\'e
Paris-Saclay, 91405 Orsay, France\\
\it $^b$ School of Physics and Astronomy, University of Southampton,\\
\it Highfield, Southampton SO17 1BJ, UK }

\end{doublespace}

\end{center}
\vspace*{2cm}
\begin{abstract}
The light higgsino-singlino scenario of the NMSSM allows to combine a naturally small
$\mu$ parameter with a good dark matter relic density. Given the new constraints
on spin-dependent and spin-independent direct detection cross sections in 2016 we study first
which regions in the plane of chargino- and LSP-masses below 300~GeV remain viable. Subsequently
we investigate the impact of searches for charginos and neutralinos at the LHC,
and find that the limits from run~I do not rule out any additional region
in this plane. Only the HL-LHC at 3000~fb$^{-1}$ will test parts of this plane
corresponding to higgsino-like charginos heavier than 150~GeV and relatively light
singlinos, but notably the most natural regions with lighter charginos seem to remain
unexplored.
\end{abstract}
\end{titlepage}

\newpage
\section{Introduction}
\label{sec:intro}

Despite considerable efforts the ATLAS and CMS collaborations have not
(yet?) discovered supersymmetric particles at the LHC \cite{atlas,cms}.
The most stringent limits are on the masses of squarks and gluinos
with strong interactions, but electroweakly interacting supersymmetric particles
(charginos $\chi^\pm_i$ and neutralinos $\chi^0_i$) can and have also been searched for.

Among the electroweakly interacting supersymmetric particles are the
higgsinos, the fermionic superpartners of the two scalar Higgs doublets
$H_u$ and $H_d$ required in supersymmetry (Susy). The higgsinos include
a chargino with a Dirac mass $\mu$, and two neutral Majorana fermions
which are part of the neutralinos. Charginos have been searched for at LEP,
and masses below $\sim 100$~GeV have been ruled out \cite{lep} implying a
corresponding lower bound on $|\mu|$.

$\mu$ is a supersymmetric mass term and generates a positive mass squared term
$\mu^2$ in the scalar potential for both SU(2) doublets $H_u$ and $H_d$, both of which must
have non-vanishing vacuum expectation values
(vevs) in order to generate up- and down-type quark masses.
At least for $H_u$, the positive mass squared term $\mu^2$ must be cancelled
by a negative soft SUSY breaking mass term $m_{H_u}^2$ such that
$\mu^2+m_{H_u}^2 < 0$ leading to $-\mu^2-m_{H_u}^2\approx M_Z^2/2$.
Such a cancellation becomes unnatural if each of these terms were
much larger than $M_Z^2$, i.e.~$\gsim (300 \text{GeV})^2$. This argument applies both in the MSSM
\cite{Haber:1984rc,Martin:1997ns,Aitchison:2005cf}
(see e.g. \cite{Chan:1997bi,Baer:2013vpa,Baer:2014kya,Bae:2015jea}) and in the NMSSM
\cite{Maniatis:2009re,Ellwanger:2009dp} (see e.g. \cite{Ellwanger:2011mu,Perelstein:2012qg}),
where $\mu\equiv \lambda \left<S\right>$ is generated by the
vev $\left<S\right>$ of a scalar singlet $S$ and $\lambda$ is a higgsino-$S$ Yukawa
coupling.

The lightest neutralino, stable if $R$-parity is conserved and if it is the lightest
supersymmetric particle (LSP), is a candidate for
dark matter. Besides the higgsinos, 
additional neutralinos are neutral electroweak gauginos (bino
and the neutral wino) and, in the NMSSM, the singlino, the supersymmetric partner
of the scalar $S$. If one assumes Grand Unification of the gaugino masses
lower bounds on the gluino mass $M_3$ imply also
lower bounds on the bino and wino masses $M_1$ and $M_2$, respectively. This assumption implies
$M_1 : M_2 : M_3 \approx 1 : 2 : 6$ depending somewhat on threshold corrections
to the running masses and radiative corrections to pole masses.
Then, if $M_3 \gsim 2$~TeV (or more) is confirmed in the
future, it is natural for the wino and bino masses $M_2, M_1$ to be larger than the
higgsino mass parameter $\mu$.

In the MSSM, the neutral higgsinos (nearly mass degenerate with the higgsino-like
charginos) would then play the r\^ole of the LSP.
However, a nearly pure higgsino LSP is known to be an imperfect candidate for
dark matter: unless very heavy with a mass beyond 1~TeV, it annihilates too fast in the early
universe implying a too small relic density \cite{ArkaniHamed:2006mb}.
(The situation might be remedied introducing additional axions as the dominant
component of dark matter in the MSSM \cite{Baer:2013vpa,Bae:2015jea}. Alternatively,
a non-thermal higgsino production in the early universe can be invoked; this
scenario is severly constrained by limits from direct dark matter detection
experiments \cite{Baer:2016ucr}.)

In the NMSSM it is natural for the singlino to be quite light; the singlino
(Majorana) mass parameter $2\kappa \left<S\right>$ (where $\kappa$ is a singlino-Singlet
Yukawa coupling) might even vanish for $\kappa\to 0$ in which case a small
singlino mass results only from mixing with the neutral higgsinos. Still, such
a mostly singlino-like LSP is good dark matter candidate \cite{Belanger:2005kh}
since its relic density can be reduced to the observed value via various processes
like a mostly singlet-like pseudoscalar in the s-channel, 
chargino exchange in the t-channel or co-annihilation. Furthermore a mostly singlino-like LSP
can easily satisfy limits from direct and indirect dark matter detection experiments;
its spin-independent direct detection cross section can even fall below the
neutrino floor \cite{Ellwanger:2014dfa,Badziak:2015exr}.

It is the purpose of the present paper to investigate in how far this attractive
scenario within the NMSSM is tested by the most recent direct dark
matter detection experiments
PICO-2L \cite{Amole:2016pye}, LUX \cite{Akerib:2016lao,Akerib:2016vxi} and
PandaX-II \cite{Tan:2016zwf,Fu:2016ega}, in combination with searches at the LHC for
electroweakinos at run~I and, in the future, by the high luminosity (HL) LHC.

At the LHC, charginos (higgsino-like or wino-like) can be pair produced, or
in association with
neutralinos $\chi^0_i$. The lightest chargino can decay via possibly virtual $W^{\pm (*)}$ bosons or, if
kinematically possible, via sleptons. The search channels are typically
missing transverse energy $E_T^\text{miss}$ plus two leptons (from
leptonic decays of two $W^\pm$ bosons in the case of chargino pair production),
or three leptons from $pp \to W^{(*)} \to \chi^\pm_1 + \chi^0_2$
with $\chi^\pm_1 \to W^{(*)} + \chi^0_1$, $\chi^0_2 \to Z^{(*)} + \chi^0_1$
and leptonic decays of both $W^{(*)}$ and $Z^{(*)}$. This $E_T^\text{miss}$
+ three lepton channel has typically the farthest reach in mass.
Another search channel is $WH$ (where $H$ is the Higgs boson of
the Standard Model), where one assumes
a $\chi^0_2$ decay of the form $\chi^0_2 \to H + \chi^0_1$, still
concentrating on the leptonic $W^{(*)}$ decay.

Previous studies of constraints on, phenomenological aspects and future prospects of
the light singlino scenario in the NMSSM have been
performed in \cite{Cerdeno:2004xw,Cerdeno:2007sn,Barger:2007nv,Belanger:2008nt,Vasquez:2010ru,
Das:2012rr,Christensen:2013dra,DaSilva:2013jga,
Perelstein:2012qg,Kozaczuk:2013spa,Cao:2013mqa,Ellwanger:2013rsa,Kim:2014noa,
Han:2014nba,Cheung:2014lqa,Dutta:2014hma,Ellwanger:2014hia,Huang:2014cla,Cahill-Rowley:2014ora,
Guo:2014gra,Cao:2014efa,Bi:2015qva,Cao:2015loa,Butter:2015fqa,Allanach:2015cia,
Gherghetta:2015ysa,
Han:2015zba,Potter:2015wsa,Barducci:2015zna,Enberg:2015qwa,Kim:2015dpa,
Badziak:2015exr,Xiang:2016ndq,Cao:2016nix,Cao:2016cnv}.
Here we proceed as follows: In the plane defined by the chargino (charged higgsino)
mass $M_{\chi^\pm_1} < 300$~GeV and the (mostly singlino-like) LSP mass
$M_{\chi^0_1} < M_{\chi^\pm_1}$
we look for regions which are or will be definitely excluded by present
dark matter searches and/or searches at the LHC. To this end we use the code 
{\sf NMSSMTools\_5.0.1}~\cite{Ellwanger:2004xm,Ellwanger:2005dv}
with NMSDECAY~\cite{Das:2011dg} to compute the spectrum, couplings and
branching fractions,
and fix the less relevant soft Susy breaking parameters to quite
pessimistic values: the squark masses
of the first two generations to 3~TeV, the slepton masses to 1~TeV
(disregarding a possible contribution to the muon anomalous magnetic moment), the
gaugino masses $M_3$, $M_2$ and $M_1$ to 2.1~TeV, 700~GeV and 350~GeV, respectively,
only $A_{top}$ and the stop masses are varied to ensure a Standard Model-like
Higgs scalar with a mass of about $125$~GeV. The NMSSM specific Yukawa
couplings $\lambda$ and $\kappa$, the soft Susy breaking terms $A_\lambda$
and $A_\kappa$, $\mu$ and $\tan\beta$ are varied for given points in
the $M_{\chi^\pm_1}-M_{\chi^0_1}$ plane. Actually $M_{\chi^\pm_1}$ defines essentially $\mu$,
and $M_{\chi^0_1}$ the ratio~$\kappa/\lambda$.

The remaining parameters are still varied in order to find the strict
boundaries in the above defined plane. These originate from a dark matter relic density
consistent with the WMAP/Planck value $\Omega_{DM}h^2=0.1187$
\cite{Hinshaw:2012aka,Ade:2013zuv}, allowing for a 10\% theory error,
and from upper limits on spin-dependent and spin-independent
direct detection cross sections from PICO-2L \cite{Amole:2016pye},
LUX \cite{Akerib:2016lao,Akerib:2016vxi} and PandaX-II \cite{Tan:2016zwf,Fu:2016ega}.
The dark matter relic density and the spin-dependent and spin-independent
direct detection cross sections are computed with help of
{\sf micr\-OMEGAS\_3}~\cite{Belanger:2013oya}.

In addition various other phenomenological constraints are required to
be satisfied. These are notably constraints from LEP on the invisible $Z$
width, from searches for charginos and associate production of neutralinos by
DELPHI \cite{Abdallah:2003xe}
and OPAL \cite{Abbiendi:2003sc}, and from searches for MSSM Higgs bosons \cite{Schael:2006cr}
(which are relevant for a lighter mostly singlet-like scalar).
Relevant constraints from the LHC stem from the ATLAS and CMS combinations
of Standard Model-like Higgs boson production and decay rates
\cite{Khachatryan:2016vau}, which constrain possible exotic
decays into pairs of NMSSM-like scalars, pseudoscalars or singlinos
with masses below $\sim 60$~GeV. The lower limit on the combined signal rates
into the $WW/ZZ$ channels is actually more relevant than the direct searches
for exotic decays.
All these constraints are implemented in {\sf NMSSMTools\_5.0.1}.

In the next section we describe the model and, in more detail, the constraints
from dark matter in the various regions of the $M_{\chi^0_1} - M_{\chi^\pm_1}$
plane. In section~3 we describe our simulations of signals at the LHC, their
comparisons with constraints from run~I and prospects for the High
Luminosity LHC (based essentially on CheckMATE \cite{Drees:2013wra,Kim:2015wza,Dercks:2016npn}).
We conclude in section~4 with a summary and outlook.

\section{The NMSSM and the singlino relic density}

We consider the NMSSM with the scale invariant superpotential
\beq\label{eq:2.1}
W_\text{NMSSM} = \lambda \hat S \hat H_u\cdot \hat H_d + \frac{\kappa}{3} 
\hat S^3 +\dots
\eeq
where the dots denote the Yukawa couplings of the superfields $\hat H_u$ and $\hat H_d$
to the quarks and leptons as in the MSSM. Once the scalar component of the superfield
$\hat S$ develops a vev $\left< S\right>\equiv s$, the first term in
$W_\text{NMSSM}$ generates an effective $\mu$-term with
\beq\label{eq:2.2}
\mu_\mathrm{eff}=\lambda\, s\; .
\eeq
Subsequently the index $_\mathrm{eff}$ of $\mu$ will be omitted for
simplicity. $\mu$ generates a Dirac mass term for the charged and
neutral SU(2) doublet higgsinos $\psi_u$ and $\psi_d$.
Additional charginos are the charged winos with a mass term $M_2$.
As stated in the introduction we assume $M_2=700$~GeV which leads to
little mixing $\lsim 3\%$ between the charged higgsinos and winos.

Of particular importance will be the neutralino sector.
Altogether the symmetric $5 \times 5$ mass matrix ${\cal M}_0$
in the basis $\psi^0 = (-i\widetilde{B} ,
-i\widetilde{W}^3, \psi_d^0, \psi_u^0, \psi_S)$
is given by \cite{Ellwanger:2009dp}
\beq\label{eq:2.3}
{\cal M}_0 =
\left( \ba{ccccc}
M_1 & 0 & -\frac{g_1 v_d}{\sqrt{2}} & \frac{g_1 v_u}{\sqrt{2}} & 0 \\
& M_2 & \frac{g_2 v_d}{\sqrt{2}} & -\frac{g_2 v_u}{\sqrt{2}} & 0 \\
& & 0 & -\mu & -\lambda v_u \\
& & & 0 & -\lambda v_d \\
& & & & 2\kappa s
\ea \right)
\eeq
where $v_u^2 + v_d^2 =v^2 \simeq (174\ \text{GeV})^2$ and $\frac{v_u}{v_d}=\tan\beta$.
The eigenstates of ${\cal M}_0$ are denoted by $\chi_i^0$, $i=1...5$ ordered in mass.
Henceforth the LSP is identified with $\chi_1^0$.

Another important r\^ole will be played by the singlet-like scalar and pseudoscalar
Higgs masses. The CP-even sector comprises three physical states which are linear
combinations of the real components $(H_{dR}, H_{uR}, S_R)$.
The (3,3) element of the $3 \times 3$ CP-even mass matrix ${\cal
M}_S^2$ reads in this basis
\beq\label{eq:2.4}
{\cal M}_{S,33}^2 \equiv M_{S_R,S_R}^2  =  \lambda A_\lambda \frac{v_u v_d}{s}
+ \kappa s (A_\kappa + 4\kappa s)\; ;
\eeq
it corresponds essentially to the mass squared of the mostly singlet-like eigenstate.
Another eigenstate must correspond to a Standard~Model-like Higgs boson $H_{125}$ with its mass
$\sim 125$~GeV, and nearly Standard~Model-like couplings to quarks, leptons and gauge
bosons. A third MSSM-like eigenstate has a mass of about
$2\frac{\mu(A_\lambda+\kappa s)}{\sin 2\beta}$. In the regions of the parameter space
of interest here we always find
that the mostly singlet-like eigenstate is the lightest CP-even scalar $H_1$, the
Standard-Model-like Higgs boson $H_{125}$ is the second lightest CP-even scalar $H_2$, and
the MSSM-like state is the third CP-even scalar $H_3$.

The CP-odd sector consists in the linear combinations of the imaginary components
$(H_{dI}, H_{uI}, S_I)$. The (3,3) element of the $3 \times 3$ CP-odd mass matrix ${\cal
M}_P^2$ reads in this basis
\beq\label{eq:2.5}
{\cal M}_{P,33}^2 \equiv M_{S_I,S_I}^2  =  \lambda (A_\lambda+4\kappa s)\frac{v_u
v_d}{s} -3\kappa A_\kappa s\; ;
\eeq
again it corresponds essentially to the mass squared of the mostly singlet-like eigenstate.
Other eigenstates are the electroweak Goldstone boson, and an MSSM-like eigenstate again with a mass of
about $2\frac{\mu(A_\lambda+\kappa s)}{\sin 2\beta}$. The masses
of the MSSM-like Higgs bosons are bounded from below to $\gsim 300-400$~GeV due to constraints
from $b\to s \gamma$ on the charged Higgs boson whose mass is similar to the ones of the CP-even and CP-odd
neutral scalars. Subsequently the lightest mostly singlet-like CP-odd eigenstate will be denoted by $A_1$.

From eqs.~\eqref{eq:2.3}-\eqref{eq:2.5} one can derive the sum rule \cite{Das:2012rr}
\beq\label{eq:2.6}
M_{\psi_S,\psi_S}^2\equiv 4\kappa^2 s^2 = M_{S_R,S_R}^2 + \frac{1}{3} M_{S_I,S_I}^2 -\frac{4}{3}
v_u v_d\left(\lambda^2\frac{A_\lambda}{\mu}+\kappa\right)
\eeq
which relates, up to modifications by mixing, the singlet-like neutralino, CP-even and CP-odd Higgs
masses. Notably for sizeable $\tan\beta$ (i.e. small $v_d$) and not too large $A_\lambda$ and Yukawa
couplings $\lambda$ and $\kappa$ the last term in eq.~\eqref{eq:2.6} is negligible.

In the following we consider, for reasons of naturalness as discussed in the introduction,
a $\mu$ parameter $\lsim 300$~GeV - below the bino and wino masses - and hence a dominantly
singlino-like LSP $\chi_1^0$ (singlino for short). Assuming a standard thermal
history of the universe with a singlino in thermal equilibrium at temperatures above
its mass, its annihilation rate must be sufficiently large such that its relic
density today complies with the WMAP/Planck value $\Omega_{DM}h^2=0.1187$
\cite{Hinshaw:2012aka,Ade:2013zuv}. Various processes can give a large enough
annihilation cross section:
\begin{itemize}
\item Annihilation via a pseudoscalar in the s-channel. For singlino masses $M_{\chi_1^0}$ below
$\mu$ as assumed here this pseudoscalar is the singlet-like $A_1$
with its mass given in eq.~\eqref{eq:2.5} (up to a small shift through mixing).
For $M_{\chi_1^0}$ below $\approx 100$~GeV, $M_{A_1}$ should be about $2\times M_{\chi_1^0}$ such
that the annihilation cross section is enhanced by the s-channel pole (depending on $\kappa$
and the mixing of $A_1$ with the MSSM-like SU(2)-doublet pseudoscalar which induces its couplings
to quarks and leptons).
For $M_{\chi_1^0}$ above $\approx 100$~GeV $M_{A_1}$ can be smaller than $2\times M_{\chi_1^0}$
allowing for the LSP annihilation via $A_1^* \to A_1+H_1$ provided $M_{H_1}$ is small enough.
For $M_{\chi_1^0}$ above $\approx m_{top}$ the annihilation via $A_1^* \to t\bar{t}$ becomes
possible.
\item Annihilation into a pair of $W$ bosons via (higgsino-like) chargino exchange
in the t-channel. This t-channel process is strong enough to be dominant only for singlino
masses above $\sim~100$~GeV.
\item Annihilation via the $Z$ boson in the s-channel if the singlino mass is about half the $Z$ mass.
\item Annihilation via the Standard~Model-like Higgs boson $H_{125}$ in the s-channel if the
singlino mass is about half the Higgs mass.
\item Coannihilation with higgsinos is possible in principle, but strongly limited
by the constraints discussed in the introduction and below.
\end{itemize}

Annihilation via $A_1 \sim S_I$ in the s-channel with a pseudoscalar mass about twice
the mass of the singlino is constrained by eq.~\eqref{eq:2.6}:
Replacing $M_{S_I,S_I}^2$ by $\sim 4M_{\psi_S,\psi_S}^2$ eq.~\eqref{eq:2.6} becomes
\beq\label{eq:2.7}
\frac{1}{3} M_{\psi_S,\psi_S}^2 = -M_{S_R,S_R}^2+\frac{4}{3}
v_u v_d\left(\lambda^2\frac{A_\lambda}{\mu}+\kappa\right)
\eeq
leading to a negative CP-even scalar mass squared if the terms $\sim v_u v_d$ are neglected.
This conclusion is avoided if one takes into account that the physical masses are obtained
through diagonalization of the mass matrices and shifted by mixing. In the cases of the
scalars this effect would be contraproductive: For small $M_{S_R,S_R}^2$ below $(125~\text{GeV})^2$,
mixing with the Standard-Model-like or MSSM-like Higgs bosons would decrease the
eigenvalue of ${\cal M}_S^2$ corresponding to $M_{H_1}^2$ even further. Likewise, mixing in the CP-odd
sector among the singlet-like and the heavier MSSM-like states reduces the physical mass
of the mostly singlet-like state $A_1$, aggravating the consequences of the sum rule \eqref{eq:2.6}
if $M_{A_1}$ is required to be about twice as large as the singlino mass.

In the neutralino sector, on the other hand, mixing reduces the mass of the singlino and
allows it to be about half of the pseudoscalar mass for positive (albeit small) CP-even
masses squared. Such mixing requires the term $-\lambda v_u$ in the mass matrix \eqref{eq:2.6}
to be not too small. Altogether the scenario where annihilation via $A_1$ with $M_{A_1}\sim 2 M_{\chi_1^0}$
leads to a satisfactory relic density is characterized by\nl
-- a light singlet-like CP-even scalar $H_1$,\nl
-- a non-negligible higgsino component of the dominantly singlino-like LSP $\chi_1^0$,\nl
-- the latter requires a non-negligible value of $\lambda\sim 0.1-0.4$ (larger for larger higgsino
masses $\sim \mu$ relative to $M_{\chi_1^0}$).

Correspondingly this scenario is subject to the following constraints:\nl
-- $H_1$ must satisfy constraints from Higgs searches at LEP,\nl
-- the $BR(H_{125}\to H_1 H_1)$ must remain below $\sim 10\%$ in order not to reduce the
Standard Model-like  signal rates of $H_{125}$ below its present limits,\nl
-- the spin-independent direct detection rate mediated by $H_1$ in the t-channel must satisfy the
($M_{\chi_1^0}$ dependent) bounds from LUX~\cite{Akerib:2016vxi} and PandaX-II \cite{Tan:2016zwf},\nl
-- for $M_{\chi_1^0}+\mu$ below $\sim 200$~GeV limits from DELPHI \cite{Abdallah:2003xe}
and OPAL \cite{Abbiendi:2003sc} on $Z^*\to \chi_1^0+\chi_{2,3}^0$ must be satisfied,\nl
-- and, of course, the relic density must coincide with the WMAP/Planck value $\Omega_{DM}
 = 0.1187 \pm 10\%$.

As stated in the introduction, the NMSSM specific Yukawa couplings $\lambda$ and $\kappa$, the soft
Susy breaking terms $A_\lambda$ and $A_\kappa$, $\mu$ and $\tan\beta$ are varied in order to
satisfy all constraints if possible. Still,
the combination of these constraints rules out regions in the $M_{\chi_1^0} - M_{\chi_1^\pm}$ plane
characterized by $M_{\chi_1^\pm}\sim \mu \lsim 240$~GeV and $M_{\chi_1^0} \lsim 100$~GeV
(for $M_{\chi_1^0} \gsim 100$~GeV, annihilation via $A_1^* \to A_1+H_1$ dominates). These
forbidden regions are indicated in blue in Fig.~1. The dominant constraints inside these blue
regions are indicated by ``relic density too large or $\sigma_{SI}$ too large'', where $\sigma_{SI}$
indicates the spin independent direct detection cross section. The limits of LUX and PandaX-II
become weaker for smaller $M_{\chi_1^0}$; for $M_{\chi_1^0}\lsim 10$~GeV the constraints from
LEP become dominant, but for $M_{\chi_1^0}\lsim 5$~GeV all constraints can be satisfied for
any value of $M_{\chi_1^\pm}$.
Of course, for $M_{\chi_1^0}\sim M_Z/2$ or $M_{\chi_1^0}\sim M_{H_{125}}/2$ annihilation via
these bosons in the s-channel is possible, and some of the above constraints are relaxed.

Another forbidden region appears for $M_{\chi_1^0}$ too close to $M_{\chi_1^\pm}\sim \mu \sim
M_{\chi_{2,3}^0}$. There mixing between $\chi_1^0$ and the higgsinos $\chi_{2,3}^0$ cannot be
avoided, and the higgsino component of $\chi_1^0$ implies a spin-dependent direct detection
rate via $Z$ exchange in conflict with constraints from LUX \cite{Akerib:2016lao} and PandaX-II
\cite{Fu:2016ega}.
(The importance of recent constraints from spin-dependent direct detection experiments on the
viable parameter space of the NMSSM has recently been underlined in \cite{Cao:2016cnv}.)
Also the relic density tends to be too small if $\chi_1^0$ has a large higgsino component.
This region is indicated by ``relic density too small and/or $\sigma_{SI}$ too large''
in Fig.~1.

\begin{figure}[t!]
\begin{center}
\includegraphics[scale=1.2,trim=0mm 0mm 0mm 0mm,clip]{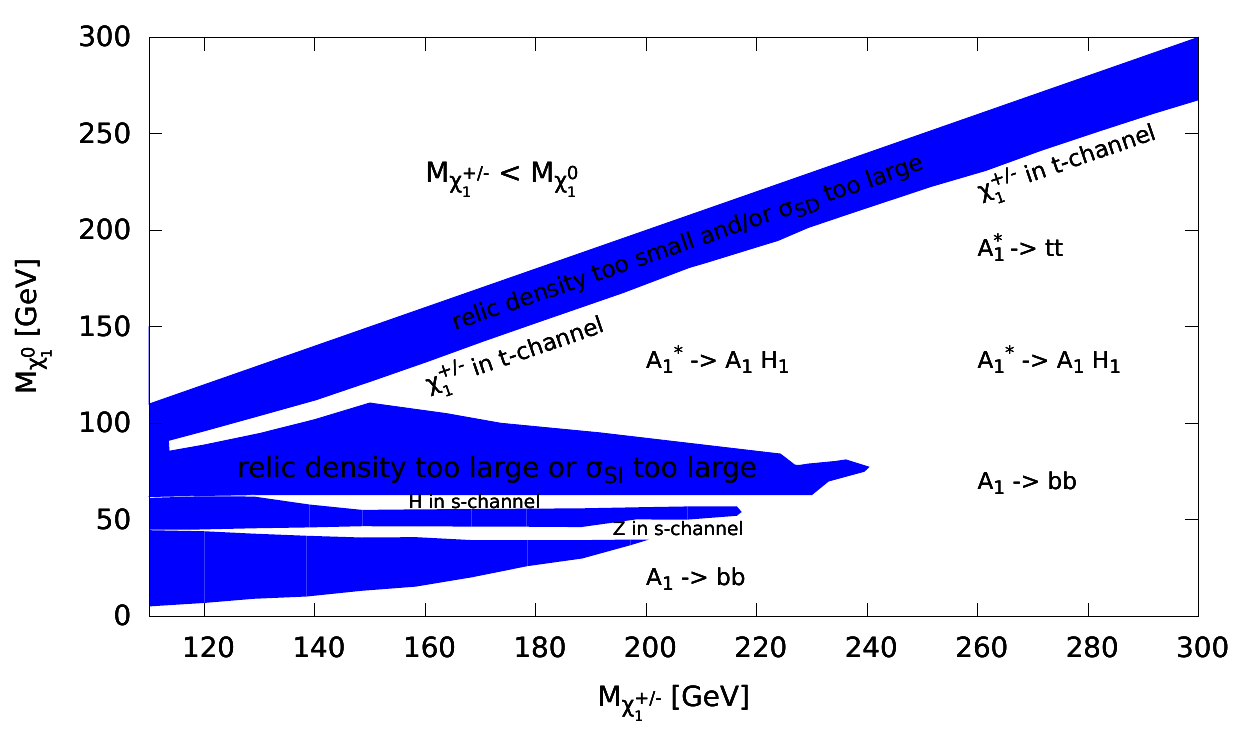}
\end{center}
\caption{In the blue regions it is impossible to satisfy simultaneously all constraints
from a good dark matter relic density, direct dark matter detection cross sections, LEP
searches for lighter Higgs bosons and neutralinos, and Higgs signal rates (i.e. the absence
of exotic decays) as measured at the LHC. In the viable white regions the dominant processes
are indicated which lead to an acceptable relic density.}
\end{figure}

In the remaining white (allowed) regions in Fig.~1 we have indicated the dominant processes
which allow for a satisfactory relic density of the singlino-like $\chi_1^0$. Of course,
not all parameter values leading to $M_{\chi_1^0}$ and $M_{\chi_1^\pm}$ in the white regions
are allowed, but at least one particular choice of parameter values gives a conflict-free
good relic density. 

Actually, due to the presence of the additional lighter mostly singlet-like Higgs
boson $H_1$ in this scenario, the upper bounds on the spin-independent dark matter cross section
mentioned above constrain the parameters everywhere: Often the t-channel
contribution of $H_1$ to the scattering amplitude has to be compensated at least partially
by a t-channel contribution of $H_{125}$ of opposite sign. Furthermore, for $M_{H_1} \lsim 60$~GeV,
the coupling $H_{125}-H_1-H_1$ must be small enough in order to suppress $H_{125}\to H_1+H_1$
decays as stated above, and LEP bounds from searches for a light Higgs boson must be respected.
These constraints can become relevant all over the white regions.

Additional constraints depend on the
dominant processes for the reduction of the relic density:\nl
-- $A_1\to bb$: Here $M_{A_1}$ has to be in the range $M_{A_1}\sim 2\dots 2.3\times M_{\chi_1^0}$
for a relic density below the WMAP/Planck value. Of course, fine tuning would be required in order to
obtain a relic density within the present $\sim 2\%$ accuracy of the combined  WMAP/Planck
measurements, but this cannot be blamed to the theory.\nl
-- $Z$ or $H_{125}$ in the s-channel: Here $M_{\chi_1^0}$ has to be in the range $M_{\chi_1^0}\sim 39\dots 49$~GeV 
(in the $Z$ case) or $M_{\chi_1^0}\sim 59\dots 64$~GeV (in the $H_{125}$ case)
for a relic density below the WMAP/Planck value. The couplings of $\chi_1^0$ to
$Z$ or $H_{125}$ are proportional to its higgsino component, and hence bounded by the limits
on the spin-dependent dark matter cross section.
For $M_{\chi_1^\pm} \sim M_{\chi_{2,3}^0} \lsim (200~\text{GeV}-M_{\chi_1^0})$,
constraints from DELPHI \cite{Abdallah:2003xe} and OPAL \cite{Abbiendi:2003sc} on
$Z^*\to \chi_1^0+\chi_{2,3}^0$ production require that the higgsino component of $\chi_1^0$
is very small, which implies small couplings to $Z$ and $H_{125}$. Accordingly $2 M_{\chi_1^0}$
has to be very close to the corresponding poles for $M_{\chi_1^\pm}\lsim 130$~GeV.
\nl
-- $\chi_1^\pm$ in the t-channel: As stated above, here $M_{\chi_1^0}$ is close to $M_{\chi_1^\pm}\sim \mu \sim
M_{\chi_{2,3}^0}$ leading to mixing between $\chi_1^0$ and the higgsinos $\chi_{2,3}^0$ inducing
potentially a too large spin-dependent direct detection cross section. The reduction of the mixing
requires the more tuning the more one approaches the corresponding forbidden blue region.

The satisfaction of all of these constraints requires to tune the dimensionless parameters
$\lambda$, $\kappa$ and $\tan\beta$ with a precision of $\sim 1-2\%$ relative to each other
nearly everywhere in the white regions, the required tuning of the dimensionful parameters $\mu$,
$A_\lambda$ and $A_\kappa$ of $\sim 10-30\%$ is weaker.
A notable exception is the case of a light $\chi_1^0$ with a
mass below 5~GeV where the present constraints on the dark matter detection cross sections
become weak or fade away: The remaining constraints from the relic density and from
$Z^*\to \chi_1^0+\chi_{2,3}^0$ production require only a relative tuning of $\approx 30\%$.
However, near the boundaries to the blue regions the parameters have
to be tuned more severely in order to satisfy the more and more stringent combinations
of constraints (which become impossible to satisfy within the blue regions), hence NMSSM points near
these boundaries are necessarily fine tuned.

\section{Present and future constraints from the LHC}

Searches for charginos and neutralinos at the LHC have been performed by
the ATLAS collaboration at run~I~\cite{ATLAS:2012crz,Aad:2014nua,Aad:2014vma,Aad:2014iza,Aad:2014yka,
Aad:2014hja,Aad:2015eda,Aad:2015jqa} and run~II~\cite{ATLAS:2016kjm,ATLAS:2016ety,ATLAS:2016uwq} 
and the CMS collaboration at run~I~\cite{Chatrchyan:2014aea,CMS:2013dea,CMS:2014ica,
Khachatryan:2014qwa,CMS:2015ela,Khachatryan:2014mma,CMS:2013afa,CMS:2016saj}
and run~II~\cite{CMS:2016gvu,CMS:2016zvj,CMS:2016hfs}.
We recall that the dominant search channels for charginos/neutralinos at the LHC are
searches for $2-3$ leptons and $E_T^{miss}$.
The absence of significant excesses in these channels is typically interpreted by the
ATLAS and CMS collaborations in terms of simplified models. These mostly assume, however,
wino-like charginos $\chi^\pm_1$ (and wino-like neutralinos $\chi^0_2$)
which have significantly larger production cross sections than
higgsinos. 

Specific searches for higgsinos have been performed by
ATLAS~\cite{ATLAS:2012crz,Aad:2014yka,Aad:2015eda} and CMS~\cite{CMS:2014ica} assuming
a lightest SUSY particle (LSP) in the form of a massless
gravitino~\cite{CMS:2014ica}, or the presence of a light stau~\cite{Aad:2014yka,Aad:2015eda}
or light sleptons~\cite{Aad:2015eda}.
Apart from these specific scenarios, light higgsinos are thus hardly
constrained by searches at the run~I of the LHC, see the analyses
within phenomenologically viable versions of the MSSM or NMSSM
in~\cite{Choudhury:2013jpa,Casas:2013pta,Belanger:2013pna,Ellwanger:2013rsa,
Baer:2013yha,Han:2013kza,Gori:2013ala,Schwaller:2013baa,Han:2015lma,Ajaib:2015yma,
Han:2013usa,Baer:2014cua,Han:2014kaa,
Eckel:2014dza,Martin:2014qra,Han:2014sya,Chakraborti:2015mra}.

The discovery potential for wino-like $\chi^\pm_1/\chi^0_2$ production
at the future and high luminosity LHC has been estimated by
ATLAS~\cite{ATLAS-PHYS-PUB-2014-010,ATL-PHYS-PUB-2015-032} and CMS~\cite{CMS:2015vka}
in the trilepton and $WH$ channels, and
seems quite promising at first sight with 95\%~CL exclusion contours
reaching up to a wino mass of about 800~GeV for a light LSP at
300~fb$^{-1}$~\cite{ATLAS-PHYS-PUB-2014-010,ATL-PHYS-PUB-2015-032,CMS:2015vka}.

First we have recast the run~I analyses from
\cite{Aad:2014nua,Aad:2015jqa,Aad:2014vma,Aad:2014iza} for the light higgsino-singlino
scenario of the NMSSM. (The exclusion reach of these analyses in the
$M_{\chi^0_1} - M_{\chi^\pm_1}$ plane is not superseeded by others.)
As in the previous section we varied the NMSSM parameters for many values of
$M_{\chi^0_1}$ and $M_{\chi^\pm_1}$. For the analyses
we employed private codes, CheckMATE \cite{Drees:2013wra,Kim:2015wza}
and CheckMATE~2 \cite{Dercks:2016npn}. 
The starting point was always the simulation of events using
MadGraph5\_aMC@NLO~\cite{Alwall:2014hca}. The output was given to the
detector simulation Delphes~3~\cite{deFavereau:2013fsa} or analysed directly
inside CheckMATE~2 \cite{Dercks:2016npn}.

The result of this recasting was that no region in the $M_{\chi^0_1} - M_{\chi^\pm_1}$
plane is conclusively excluded at the 95\% level by the available 3-lepton + $E_T^{miss}$
(and 1-lepton + $H_{125} +E_T^{miss}$) searches at run~I, and this not only
marginally. The reasons for this result are\nl
-- the smaller production cross section for higgsinos compared to winos;\nl
-- a further reduction of the higgsino production cross section due to mixing
of the neutral higgsinos with the singlino;\nl
-- sizeable branching fractions of the neutral higgsinos into the singlino-like
LSP + the light singlet-like scalar $H_1$ (and, less prominently, the light singlet-like
pseudoscalar $A_1$) notably if decays via $Z$ and/or $H_{125}$ are kinematically
forbidden. These decays escape detection in the $WH$ final state due to the
much smaller singlet-like Higgs mass.

Next we recast the prospects for chargino-neutralino searches at the
HL-LHC at 3000~fb$^{-1}$ into the higgsino-singlino scenario of the NMSSM.
The analyses from \cite{ATLAS-PHYS-PUB-2014-010} (which have similar exclusion/detection
reaches as the ones in \cite{ATL-PHYS-PUB-2015-032,CMS:2015vka}) are implemented
in CheckMATE~2 \cite{Dercks:2016npn}. 

The relevant signal regions are again the ones dedicated to $WZ$ and $WH_{125}$
final states, with $Z$ or $H_{125}$ originating from $\chi^0_2\to \chi^0_1+Z/H_{125}$
decays. Since the search for $WZ$ requires
two same-flavour opposite-sign leptons with an invariant mass close to $M_Z$,
it becomes sensitive to the light higgsino-singlino scenario of the NMSSM for
$M_{\chi^0_2}\sim M_{\chi^0_3}\ (\sim M_{\chi^\pm_1}) > M_{\chi^0_1} +M_Z$ which
is indicated in Fig.~2 as a black line (below which the inequality is satisfied).
Discovery of NMSSM points below this line is thus possible
for selected points, since the production cross sections and branching fractions can be large enough for
higgsino masses below 300~GeV despite their reduced couplings.

However, even a ``3~$\sigma$ discovery'' is not necessarily guaranteed,
amongst others for the same reasons as for the run~I. Another
reason are large cuts on the lepton $p_T$ and
$E_T^{miss}$, which are required to cope with the larger instantaneous luminosity
(and the larger background), see the corresponding cuts
in~\cite{ATLAS-PHYS-PUB-2014-010,ATL-PHYS-PUB-2015-032,CMS:2015vka}. 
So we ask again whether, varying all parameters within the imposed constraints,
any region in the $M_{\chi^0_1} - M_{\chi^\pm_1}$ plane can be conclusively excluded
at the 95\% level at the HL-LHC.
The cuts lead to large enough acceptances only for mass differences
$M_{\chi^\pm_1}-M_{\chi^0_1}\gsim 150\dots 200$~GeV (after suffering from the same
reductions of the production cross sections and variations of the branching fractions
as at the run~I). There the decay $\chi^0_{2,3}\to \chi^0_1 + H_{125}$ becomes also
possible with a sufficiently large acceptance.

The region in the $M_{\chi^0_1} - M_{\chi^\pm_1}$ plane which can be
excluded at the 95\% CL level is shown in red in Fig.~2.
In CheckMATE~2 \cite{Dercks:2016npn} an additional search region is proposed, which
is based on 2~leptons + $E_T^{miss}$ +$W_{had}$, the hadronic decays of $W$ from the
$\chi_1^\pm \to \chi_1^0 + W^\pm$ decays. Due to the larger branching fraction of $W_{had}$
this search may cover a slightly larger region in the $M_{\chi^0_1} - M_{\chi^\pm_1}$ plane,
which is indicated in yellow in Fig.~2. The viability of this search channel remains
to be confirmed by the experimental collaborations, however.
In any case, even the HL-LHC at 3000~fb$^{-1}$ seems unable to test the full
$M_{\chi^0_1} - M_{\chi^\pm_1}$ plane in the  light higgsino-singlino
scenario of the NMSSM.

\begin{figure}[t!]
\begin{center}
\includegraphics[scale=1.2,trim=0mm 0mm 0mm 0mm,clip]{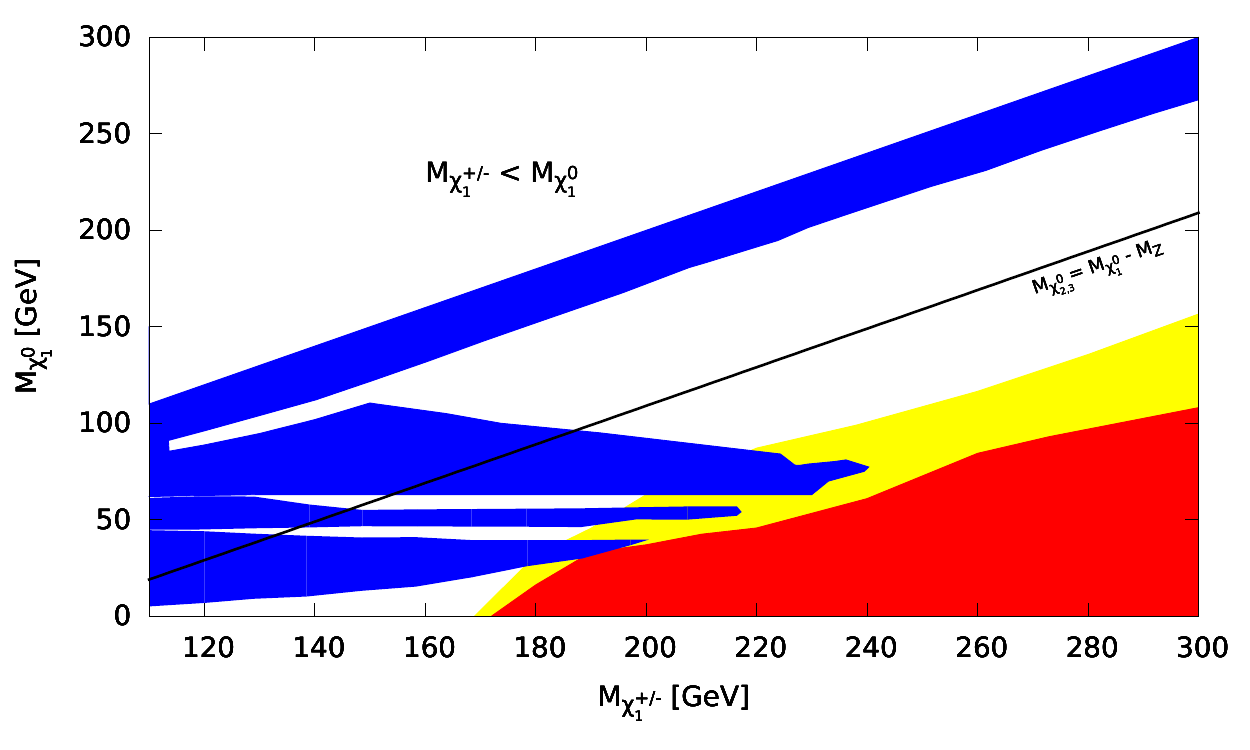}
\end{center}
\caption{The red region can be excluded at 95\% CL at the HL-LHC at 3000~fb$^{-1}$
using a recast of the prospects for chargino-neutralino searches by
ATLAS~\cite{ATLAS-PHYS-PUB-2014-010}.
The yellow region can be excluded according to a search proposed in
CheckMATE~2~\cite{Dercks:2016npn} including hadronic $W$ decays. The blue region is
the same as in~Fig.~1.}
\end{figure}

In Fig.~2 we have indicated again in blue the regions which are excluded by the
combination of constraints notably from the dark matter relic density, and the
present constraints from direct dark matter detection. In the absence of signals
in the future these latter constraints would become stronger, but we have checked
that the boundaries would hardly move. The reason is that the LSP can remain
dominantly singlino-like in most regions except for a singlino mass very close
to the higgsino (chargino) masses; hence the conclusions would hardly change.

Although one observes a certain complementarity between the constraints from dark
matter and the possible future limits at the LHC it is obvious that large
regions in the $M_{\chi^0_1} - M_{\chi^\pm_1}$ plane will remain unexplored.
This concerns in particular well motivated and natural regions where
$M_{\chi^\pm_1}\lsim 150$~GeV and $M_{\chi^0_1}\lsim 5$~GeV; such light dark
matter candidates are difficult to discover via direct detection and the hope was
that they are easier to discover at the LHC.

\section{Summary and Conclusions}

The light higgsino-singlino scenario of the NMSSM is probably the most attractive
scenario for supersymmetric dark matter, since it allows to combine a small
$\mu$ parameter with a good relic density and alleviated constraints from
dark matter searches. First we have shown that this remains so after the updated
constraints on spin-independent and spin-dependent cross sections in 2016.
Then we have studied present and future constraints on this scenario from
the LHC, and found that even the HL-LHC seems not able to test all
possible realizations of this scenario. Possible ways to improve this situation
could be:
\begin{itemize}
\item To rely on the larger production cross sections of wino-like charginos and
neutralinos, and the discovery of lighter higgsinos/singlino via cascade decays.
Of course this depends on additional parameters like the wino- (and possibly
slepton-) masses and branching fractions. Moreover the transverse momenta of the
leptons and $E_T^{miss}$ would again be reduced through multiple cascades.
\item To try to become sensitive to smaller transverse momenta of the
leptons and $E_T^{miss}$ even at 14~TeV and large instantaneous luminosity
by recording and analysing only a fraction of the events.
\end{itemize}
If such attempts turn out to be fruitless, this attractive scenario could
be tested only at an electron-positron collider~\cite{Baer:2014yta}.

\section*{Acknowledgements}

U.~E. acknowledges support from the European Union Initial
Training Network Higgs\-Tools (PITN-GA-2012-316704), from
the ERC advanced grant Higgs@LHC, from the European Union's Horizon 2020
research and innovation programme H2020-MSCA-RISE-2014 No. 645722
(NonMinimalHiggs) and under
the Marie Sklodowska-Curie grant agreements No 690575 and No 674896.


\end{document}